\documentclass[]{piparticle-final}
\usepackage{graphicx}
\usepackage{amsmath}

\begin{document}
\volume{1}               
\articlenumber{010006}   
\journalyear{2009}       
\editor{A. Go\~{n}i}   
\received{10 November 2009}     
\accepted{26 November 2009}   
\runningauthor{H. Arabshahi \itshape{et al.}}  
\doi{010006}         

\title{Light effect in photoionization of traps in GaN MESFETs}

\author{H. Arabshahi,\cite{inst1}\thanks{E-mail: arabshahi@um.ac.ir}
        A. Binesh\cite{inst2}
        }

\pipabstract{Trapping of hot electron behavior by trap centers located in buffer
layer of a wurtzite phase GaN MESFET has been simulated using an
ensemble Monte Carlo simulation. The results of the simulation show that the trap
centers are responsible for current collapse in GaN MESFET at low
temperatures. These electrical traps degrade the performance of the
device at low temperature. On the opposite, a light-induced increase in the trap-limited drain current, results from the photoionization of trapped carriers and their return to the channel under the influence of the built in electric field associated with the trapped charge distribution. The simulated device geometries and doping are matched to the nominal parameters described for the experimental structures as
closely as possible, and the predicted drain current and other
electrical characteristics for the simulated device including trapping
center effects show close agreement with the available
experimental data.} \maketitle

\blfootnote{
\begin{theaffiliation}{99}
   \institution{inst1} Department of Physics, Ferdowsi University of Mashhad, P.O. Box 91775-1436, Mashhad, Iran.
   \institution{inst2} Department of Physics, Payam-e-Nour University, Fariman, Iran.
\end{theaffiliation}
}


\section{Introduction}
GaN has become an attractive material for power transistors [1-3] due to its wide band gap, high breakdown electric
field strength, and high thermal conductivity. It also has a
relatively high electron saturation drift velocity and low relative
permitivity, implying potential for high frequency performance.
However, set against the virtues of the material are the disadvantages
associated with material quality. GaN substrates are not readily
available and the lattice mismatch of GaN to the
different substrate materials commonly used means that layers
typically contain between 10$^8$ and 10$^{10}$ threading
dislocations per cm$^2$. Further, several types of electron traps
occur in the device layers and have a significant effect on GaN devices.\\
In the search for greater power and speed
performance, the consideration of different aspects that severely limit
the output power of GaN FETs must be accounted for. It is found that
presence of trapping centers in the GaN material is the most important
phenomenon which can effect on current collapse in output drain
current of GaN MESFET. This effect was recently experimentally
investigated in GaN MESFET and was observed that the excess charge
associated with the trapped electrons produces a depletion region in
the conducting channel which results in a severe reduction in drain
current [4]. The effect can be reversed by librating trapped electrons
either thermally by emission at elevated temperatures or optically by
photoionization. There have been several experimental studies of the effect of
trapping levels on current collapse in GaN MESFET. For example, Klein
{\it et al.} [5-6] measured photoionization spectroscopy of traps in GaN
MESFET transistors and calculated that the current collapse resulted from
charge trapping in the buffer layer. Binari {\it et al.} [7] observed
decreases in the drain current of a GaN FET corresponding to the deep
trap centers located at 1.8 and 2.85 eV.\\
In this work, we report a Monte Carlo simulation which
is used to model electron transport in wurtzite GaN MESFET including a
trapping centers effect. This model is based upon the fact that since
optical effect can emit the trapped electrons that are responsible
for current collapse, the incident light wavelength dependence of this effect should
reflect the influence of trap centers on hot electron transport
properties in this device. This article is organized as
follows. Details of the device fabrication and trapping model
which is used in the simulated device are presented in section 2, and
the results from the simulation carried out on the device are
interpreted in section 3.\\

\section{Model, device and simulations}

An ensemble Monte Carlo simulation has been carried out to simulate
the electron transport properties in GaN MESFET. The method simulates
the motion of charge carriers through the device by following the
progress of $\rm 10^{4}$ superparticles. These particles
are propagated classically between collisions according to their
velocity, effective mass and the prevailing field. The selection of
the propagation time, scattering mechanism and other related quantities,
is achieved by generating random numbers and using these numbers to
select, for example, a scattering mechanism. Our self-consistent Monte
Carlo simulation was performed using an analytical band structure model
consisting of five non-parabolic ellipsoidal valleys. The scattering mechanisms 
considered for the model are acoustic and polar optical phonon, ionized impurity, piezoelectric and nonequivalent intervalley scattering. The nonequivalent intervalley scattering is between the $\Gamma_{1}$, $\Gamma_{3}$, U, M and K points.\\
The parameters used for the present Monte Carlo simulations for wurtzite GaN are the same as those used by Arabshahi for MESFET transistors [8-9].
\begin{figure}[bt]
\includegraphics[width=0.5\textwidth]{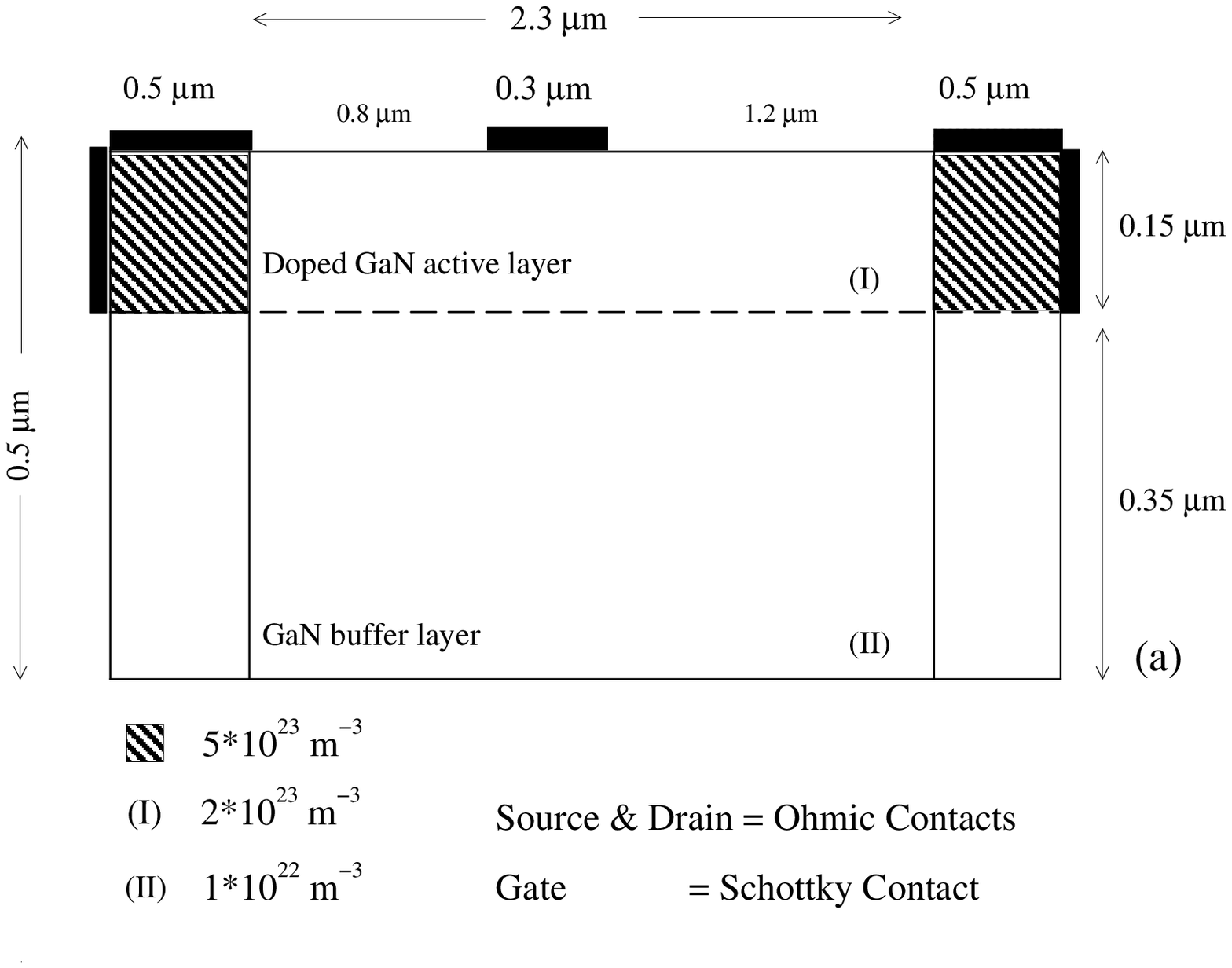}
\includegraphics[width=0.5\textwidth]{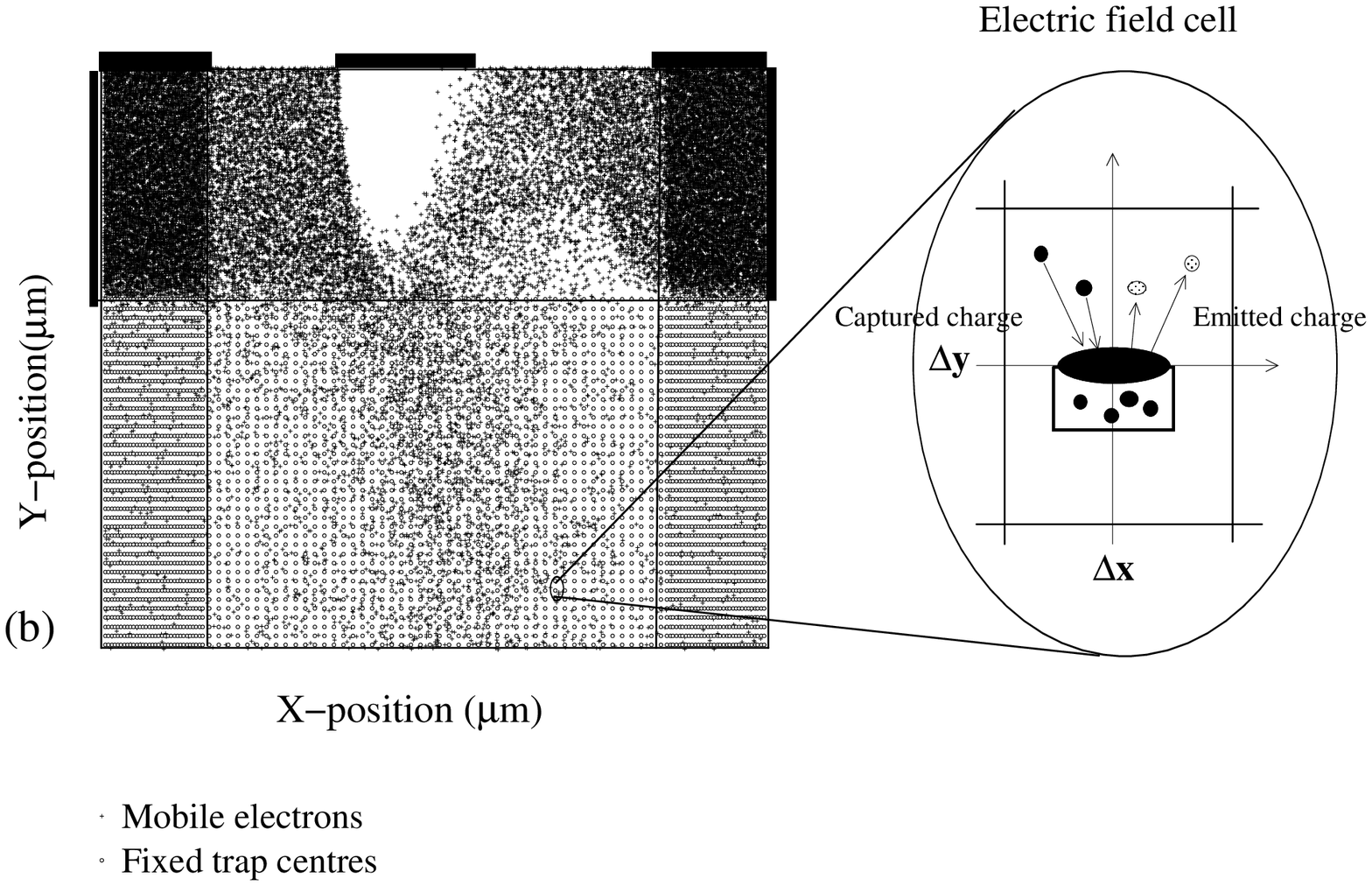}
\caption{(a) Cross section of wurtzite GaN MESFET structure which we have
chosen in our simulation. Source and drain contacts have low resistance ohmic
contacts, while the gate contact forms a Schottky barrier between the
metal and the semiconductor epilayer, (b) The instantaneous distribution of
$\rm 10^{4}$ particles at steady forward bias (drain voltage 50 V,
gate voltage $-1$ V), superimposed on the mesh. Note that in
the simulation there are two types of superparticles. The mobile particles which
describe unbound electron flow through the device and trapping center
particles which are fixed at the center of each electric field cell
(in this case in the buffer layer only). The ellipse represents a trap
center which is fixed at the center of an electric field cell and
occupied by some mobile charges.}
\end{figure}
The device structure illustrated in figure 1.a is used in all simulations. The overall device length is 3.3 $\mu$m in the $\it
x$-direction and the device has a 0.3 $\mu$m gate length and 0.5
$\mu$m source and drain length. The source and drain have ohmic
contacts and the gate is in Shottky contact in 1 eV to reperesent the
contact potential at the Au/Pt. The source and drain regions are doped
to $\rm 5\times 10^{23}$ $\rm m^{-3}$ and the top and down buffer
layers are doped to $\rm 2\times 10^{23}$ $\rm m^{-3}$ and  $\rm
1\times 10^{22}$ $\rm m^{-3}$, respectively. The effective source to
gate and gate to drain separation are 0.8 $\mu$m and 1.2 $\mu$m,
respectively. The large dimensions of the device need a long
simulation time to ensure convergence of the simulator. The device is
simulated at room temperature and 420 K.\\
In the interests of simplicity it is assumed that there is just a
single trap with associated energy level $E_{T}$ in all or just part of the
device. Further, it is assumed that only electrons may be
captured from the conduction band by the trap centers, which have a
capture cross-section $\sigma_{n}$ and are neutral
when unoccupied, and may only be emitted from an occupied center to the
conduction band. We use the standard model of carrier trapping and
emission [9-10].\\
For including trapping center effects, the following assumption has
been considered.
The superparticles in the ensemble Monte Carlo simulation are assumed
to be of two types. There are mobile particles that represent unbound
electrons throughout the device. However, the particles may also undergo
spontaneous capture by the trap centers distributed in the device. The
other type of superparticles are trapping centers that
are fixed at the center of each mesh cell. As illustrated in
figure 1.b, each trap center has the capacity to trap a finite
amount of mobile electronic charge from particles that are in its vicinity and
reside in the lowest conduction band valley. The
vicinity is defined as exactly the area covered by the electric field
mesh cell. The finite capacity of the trapping center in each cell of
a specific region in the device is set by a density parameter in the
simulation programme. The simulation itself is carried out by the
following sequence of events. First, the device is initialized with a
specific trap which is characterized by its density as a function of
position, a trap energy level and a capture cross-section. Then at a
specific gate bias the source-drain voltage is applied.\\
Some of the mobile charges passing from the source to the drain in
each timestep can be trapped by the centers with a probability which
is dependent on the trap cross-section and particle velocity in the
cell occupied at the relevant time {\it t}. The quantity of charge
that is captured from a passing mobile particle is the product of this
probability and the charge on it. This charge is deducted from the
charge of the mobile particle and added to the fixed charge of the
trap center. The emission of charge is simulated using the emission
probability. Any charge emitted from a trap center is evenly distributed
to all mobile particles in the same field cell. Such capture
and emission simulations are performed for the entire mesh in the device and
information on the ensemble of particles is recorded in the usual way.
\section{Results}
The application of a high drain-source voltage
causes hot electrons to be injected into the buffer layer where they
are trapped by trap centers. The trapped electrons produce a depletion
region in the channel of the GaN MESFET which tends to pinch off the
device and reduce the drain current. This effect can be reversed by
any factor which substantially increases the electron emission
rate from the trapped centers, such as the elevated temperatures
considered previously. Here we consider the effect of exposure to
light [11-13].\\
There have been several experimental investigations of the influence
of light on the device characteristics. Binari {\it et al.} [6] were the first to experimentally study the
current collapse in GaN MESFETs as a function of temperature and
illumination. They showed that the photoionization of trapped
electrons in the high-resistivity GaN layers and the subsequent return
of these electrons to the conduction band could reverse the drain
current collapse. Their measurements were carried out as a function of
incident light wavelength with values in the range of 380 nm to 720 nm,
corresponding to photon energies up to 3.25 eV which is close to the
GaN band gap. Their results show that when the photon
energy exceeds the trap energy, the electrons are quickly emitted and a
normal set of drain characteristics is observed.\\
To examine the photoionization effect in our simulations, the thermal
emission rate $e^{t}_{n}$ was changed to $e^{t}_{n}+e^{o}_{n}$, where
$e^{o}_{n} \sim \sigma^{o}_{n} \Phi$ is the optical emission rate,
with $\sigma^{o}_{n}$, the optical capture cross-section and $\Phi$
the photon flux density given by
\begin{equation}
\Phi=\frac{I}{h\nu}=\frac{I\lambda}{hc}
\end{equation}
where {\it I} is the light intensity, $\nu$ is the radiation frequency
and $\lambda$ is the incident light wavelength.\\
\begin{figure}[bt]
\includegraphics[width=0.45\textwidth]{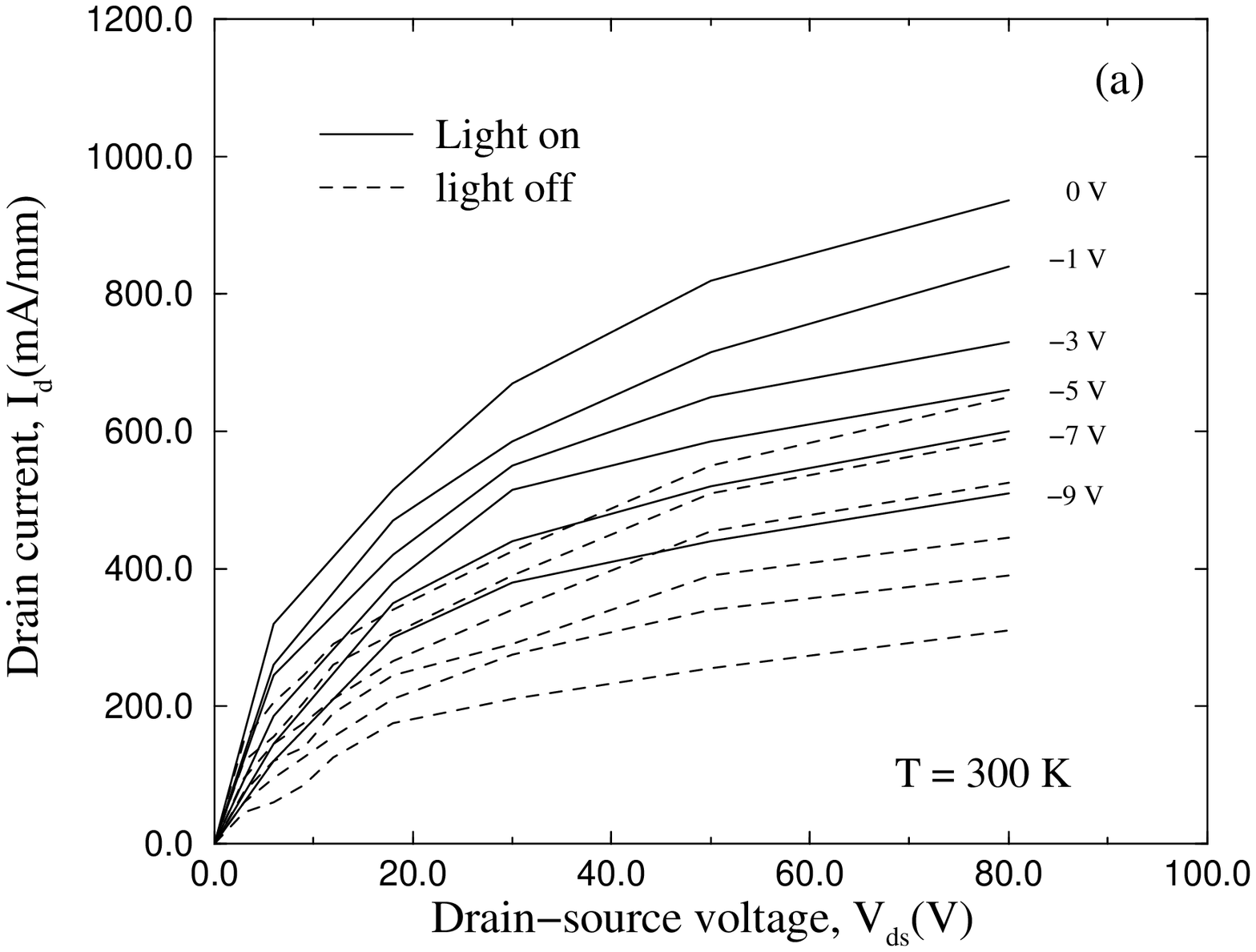}
\includegraphics[width=0.45\textwidth]{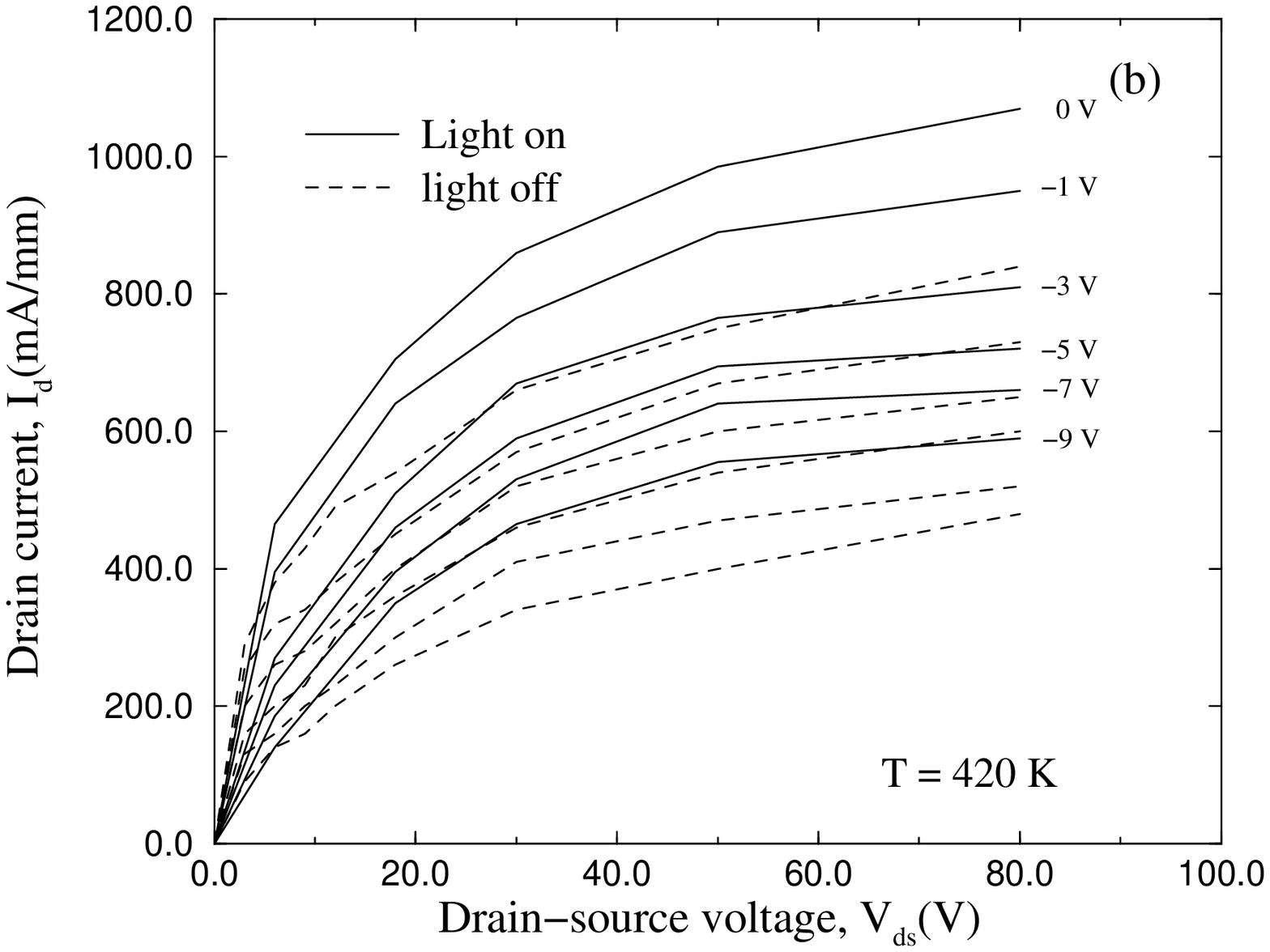}
\caption{{\it I-V} characteristics of a GaN MESFET
under optical and thermal emission of trapped electrons (solid curve)
and thermal emission of trapped electrons (dashed curve) at two different
temperatures. (a) At $T=300$ K with trap centers at 1.8 eV and
illuminated with a photon energy of 2.07 eV. (b) At $T=420$ K with
trap centers at 2.85 eV and illuminated with a photon energy of 3.1 eV.}
\end{figure}
Our modeling of photoionization effects in GaN MESFETs is based
on parameters used by Binari and Klein [5-7]. The simulations
were all carried out for two different deep trap centers, both with a
concentration of $\rm 10^{22}$ $\rm m^{-3}$, and with
photoionization threshold energies at 1.8 and 2.85 eV and capture
cross-sections of $\rm 6\times 10^{-21}$ $\rm m^{2}$ and $\rm
2.8\times 10^{-19}$ $\rm m^{2}$, respectively. A fixed incident light
intensity of 5 $\rm Wm^{-2}$ at photon energies of 2.07 eV and 3.1 eV
is used. The simulations have been performed at a sufficiently high
temperature (420 K), for both thermal and optical emission, to be
significant as well as at room temperature.\\
Figure 2a illustrates the effect on the drain current
characteristics of exposure of the device to light at room
temperature. The GaN MESFET has a deep trap center at 1.8
eV and is illuminated at a photon energy of 2.07 eV. It can be
seen that in the light the {\it I-V} curves generally exhibit a larger
drain current, especially at higher drain voltages, reflecting the
fact that the density of trapped electrons is much lower.\\
Simulations have also been performed at 420 K for a device with deep
level traps at 2.85 eV. The simulation results in figure 2b for
illumination of a photon energy of 3.1 eV are compared with the
collapsed {\it I-V} curves in the absence of light. Comparison of
figures 2a and 2b shows that the currents are generally higher
at 420 K and that the light has less effect at the highest temperature. 
\section{Conclusions}
The dependence upon light intensity (exposure) of the reversal of current collapse was simulated in a GaN MESFET for a single tapping center. Traps in the simulated device produce a serious reduction in
the drain current and consequently the output power of GaN MESFET. The
drain current behavior as a function of illumination with photon energy was also studied.
Our results show that as the temperature and photon energy are increased, the
collapsed drain current curve moves up toward the non-collapsed curve
due to more emission of trapped electrons.

\begin{acknowledgements}
The authors wish to thank M. G. Paeezi for the helpful comments and critical reading of the manuscript.
\end{acknowledgements}


\begin{thebibliography}{50}

\bibitem{Bernard} B Gil, {\it Group-III Nitride Semiconductor Compounds}, Oxford Science Pub. (1998).

\bibitem{Asif} M A Khan, M S Shur, {\it AlGaN/GaN Metal Oxide Semiconductor Heterostructure Field Effect Transistor}, Mater. Sci. Eng. {\bf B 42}, 69 (1997).

\bibitem{Binari1} P B Klein, S C Binari, J A Freitas, A E Wickenden, {\it Photoionization spectroscopy of traps in GaN metal-semiconductor field-effect transistors}, J. Appl. Phys. {\bf 88}, 2843 (2000).

\bibitem{Khan} M A Khan, M S Shur, Q C Chen, J N Kuznia, {\it Low frequency noise in GaN metal semiconductor and metal oxide semiconductor field effect transistors}, Electron. Lett. {\bf 30}, 2175 (1994).

\bibitem{Klein1} P B Klein, S C Binari, J A Freitas, A E Wickenden, {\it Observation of deep traps responsible for current collapse in GaN metal-semiconductor field-effect transistors}, J. Appl. Phys. {\bf 88}, 2843 (2000).

\bibitem{Klein2} P B Klein, J A Freitas, S C Binari, A E Wickenden, {\it AlGaN/GaN heterostructure field-effect transistor model including thermal effects}, Appl. Phys. Lett. {\bf 75}, 4016 (1999).

\bibitem{Binari2} S C Binari, W Kruppa, H B Dietrich, G Kelner, A E Wickenden, J A Freitas, {\it Trapping effects and microwave power performance in AlGaN/GaN HEMTs}, Solid State Electron. {\bf 41},
1549 (1997).

\bibitem{Arabshahi1} H Arabshahi, {\it Monte Carlo simulations of electron transport in Wurtzite phase GaN MESFET including trapping effect}, Modern Phys. Lett. B {\bf 20}, 787 (2006).

\raggedbottom
\pagebreak

\bibitem{Arabshahi2} H Arabshahi, {\it The frequency response and effect of trap parameters on the characteristic of GaN MESFETs}, The Journal of Damghan University of Basic Sciences {\bf 1},
45 (2007).

\bibitem{Trassaert} S Trassaert, B Boudart, C Gaquiere, {\it Investigation of traps induced current collapse in GaN devices}, a1404 ORSAY France, 127 (1999).

\bibitem{Kastalsky} A Kastalsky, S Luryi, A C Gossard, W K Chan, {\it Switching in NERFET circuits}, IEEE Electron Device Lett. {\bf 6}, 347 (1985).

\bibitem{Inkson} J C Inkson, {\it Deep impurities in semiconductors. II. The optical cross section}, J. Phys. C: Solid State Phys. {\bf 14}, 1093 (1981).

\bibitem{Lang} D V Lang, R A Logan, M Jaros, {\it Monte Carlo evaluations of degeneracy and interface roughness effects}, Phys. Rev. B {\bf 19}, 1015 (1979).


\end{thebibliography}
\end{document}